# Spin-Texture Spin-valve with a van der Waals Magnet

Bing Zhao[1], Roselle Ngaloy[1], Lars Sjöström[1], Saroj P. Dash[1,2,3*]

[1]*Department of Microtechnology and Nanoscience, Chalmers University of Technology, SE-41296, Göteborg, Sweden*
[2]*Wallenberg Initiative Materials Science for Sustainability, Department of Microtechnology and Nanoscience, Chalmers University of Technology, SE-41296, Göteborg, Sweden.*
[3]*Graphene Center, Chalmers University of Technology, SE-41296, Göteborg, Sweden.*

**Abstract**

All-electrical methods for nucleating, detecting, and manipulating spin textures in two-dimensional (2D) van der Waals (vdW) magnets can serve as fundamental building blocks for multi-state spintronic memory, logic, and neuromorphic computing applications. Unlike conventional ferromagnets, vdW ferromagnet such as $Fe_5GeTe_2$ with strong Dzyaloshinskii-Moriya interactions stabilize nanoscale chiral spin textures including skyrmions and stripe domains. However, the sub-100 nm size of these spin textures has limited their study to sophisticated microscopy techniques. Here, we demonstrate all-electrical detection of spin textures in vdW itinerant ferromagnet $Fe_5GeTe_2$ using pure spin transport in a lateral graphene spin-valve device at room temperature. By engineering nanoscale constrictions or notches in $Fe_5GeTe_2$, we create spin textures that inject distinct spin polarizations into the graphene channel, where they are nonlocally sensed by a reference conventional ferromagnetic detector at room temperature. This enables the observation of anomalous multi-level spin-valve switching and Hanle spin precession signals, which are due to unique spin textures in $Fe_5GeTe_2$ and in sharp contrast to single-domains and conventional magnet-based devices. This all-electrical approach can provide direct access to the spin textures on an integrated 2D spintronic circuit without the need for ex-situ microscopic characterizations.

**Introduction**

Spin textures in magnetic materials arise when a discrete symmetry is spontaneously broken, such as in domain walls or skyrmions [1–3]. In recent times, there has been significant interest in the manipulation and dynamic excitation of domain walls and skyrmions using spin-polarized currents, magnetic fields, and electric fields [4–9]. The creation, detection, movement, and control of such spin textures are crucial for data storage and retrieval, which are essential for applications in memory and logic technologies [3,7]. However, the efficient electrical detection and manipulation of spin textures by spin currents or electric fields require the development of new low-dimensional materials and heterostructures [10,11].

The discovery of van der Waals (vdW) magnets has the potential for efficient control of magnetism and spin dynamics down to atomically-thin two-dimensional (2D) layers [12,13]. Recently, vdW magnetic materials have shown great potential for tunnel magnetoresistance, spin-valves, skyrmions, and spin-orbit torque-based memory devices [14–40]. $Fe_5GeTe_2$ belongs to an emerging class of vdW magnets that host intrinsic spin textures driven by Dzyaloshinskii-Moriya interactions (DMI) attributed to the material's non-centrosymmetric superstructure —a phenomenon not observed in traditional ferromagnets. $Fe_5GeTe_2$ hosts coexisting exotic stripe magnetic domains and multiple skyrmionic spin textures in one material, offering freedom for designing novel spintronic devices [21,41–45]. The presence of a variety of mixed Bloch-Néel chiral spin textures, including stripe domain textures in vdW magnets, can result from the interplay between magnetism, spin-orbit coupling, and broken symmetry of crystals that can give rise to the interplay between the dipolar and the DMI[41,44,46–48]. However, the characterization of spin textures in such vdW magnets is so far mainly limited to microscopy techniques [49] such as the magneto-optical Kerr effect [14,15], nitrogen-vacancy magnetometry [50], magnetic force microscopy [21,26], X-ray photoemission electron microscopy [51], and Lorentz transmission electron microscopy [52]. The development of electrical techniques for detecting spin textures in such vdW magnets is of great interest for next-generation information storage and processing technologies.

Here, we demonstrate for the first time the electrical detection of the magnetic spin texture in a vdW magnet $Fe_5GeTe_2$ using pure spin current in a graphene non-local spin-valve device at room temperature. To create nucleation sites for spin textures, we utilized $Fe_5GeTe_2$ nanolayers with constrictions and notches. Unlike conventional ferromagnets where constrictions create purely geometric domain patterns, strong DMI in $Fe_5GeTe_2$ can enable creation of spin textures. Such a $Fe_5GeTe_2$ nanolayer with spin texture, enabled the injection of different spin orientations into the graphene channel, resulting in anomalous multi-level spin-valve switching and corresponding Hanle spin precession signals. Such a signature of spin-textured spin-valves is observed for the first time, which is drastically different from the regular-shaped $Fe_5GeTe_2$ with single domains. In comparison, one cannot even obtain such a spin texture in traditional ferromagnets, solely through geometrical constructions/notches due to their low DMI. These findings provide important insights into the creation and detection of magnetic spin textures in $Fe_5GeTe_2$ using electrical methods in vdW-heterostructure-based graphene spintronic devices.

**Results and Discussion**

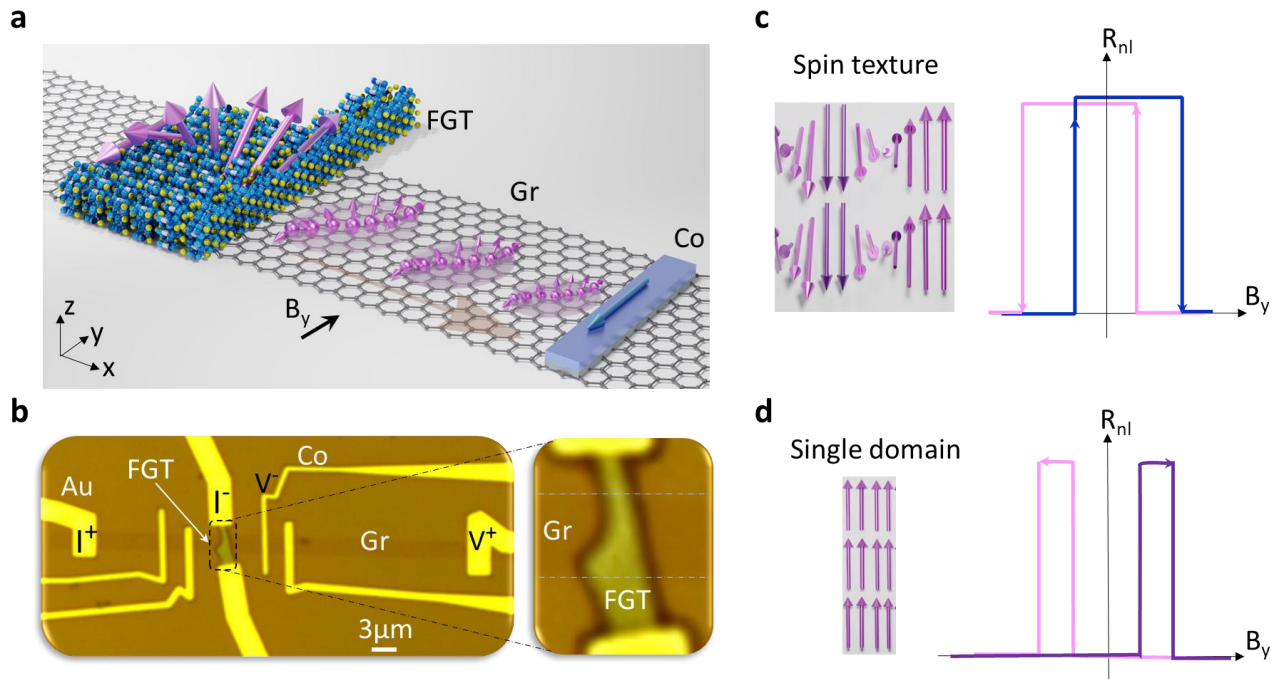

*Figure 1. Concept of electrical detection of spin textures in van der Waals magnetic spin-valve. a, b.* Schematics and optical image of Dev 1 consisting of FGT with a constriction to host magnetic spin textures, which is used for spin injection into the CVD graphene channel with reference magnetic (Co) and nonmagnetic (Au) contacts. The arrows in a) represent the spin textures and spins injected into the graphene channel. $B_y$ is the magnetic field along the y-axis. The right panel of b) shows a magnified microscope picture of the FGT flake with a constriction. Current $I^{(-/+)}$ and voltage $V^{(+/-)}$ are the electrical connections for the nonlocal spin-valve measurement geometry. *c, d.* Illustration of the spin textures (ST) and conventional single domain (SD), and the corresponding spin valve signals.

We investigate spin texture formation in the vdW $Fe_5GeTe_2$, which can have DMI one to two orders of magnitude stronger than conventional ferromagnets. While geometric constrictions in traditional ferromagnets create simple domain-wall pinning sites, the same constrictions in $Fe_5GeTe_2$ with strong DMI can nucleate complex chiral spin configurations including skyrmions and stripe domains. The schematic diagram of the lateral spin-valve device with a constricted Fe$_5$GeTe$_2$ (FGT) electrode is shown in Fig. 1a, with the corresponding optical image and the nonlocal measurement geometry of Dev 1 in Fig. 1b (see details of the device fabrication in the Methods section). Here, we use a spin valve device having one electrode with spin-texture (ST) in vdW ferromagnet FGT and another conventional electrode (Co) with single-domain (SD) in a non-local spin valve device. The vdW magnet FGT is known for the non-centrosymmetric √3×√3 superstructures [23,41] and exotic chiral spin structures, like stripe magnetic domain patterns and skyrmionic spin textures, which is attributed to the DMI[44,47]; however, they can be randomly generated or arranged in the layers. Therefore, to create skyrmion nucleation sites, structural patterning of the magnetic film with constrictions, notches, or defects is commonly used in conventional ferromagnetic materials [53–56]. Here, we utilized a FGT nanolayer with a constriction or a notch for the formation and pinning of spin texture at the FGT/graphene interface (Fig. 1b). The spins are then injected from the FGT spin texture into the graphene channel, where spins are transported and finally detected by a standard reference cobalt (Co) electrode with a quasi-single domain. The well-known magnetic switching properties of the reference Co detector electrode in a graphene spin-valve devices [57,58] allow the detection of new spin textures of the FGT electrode.

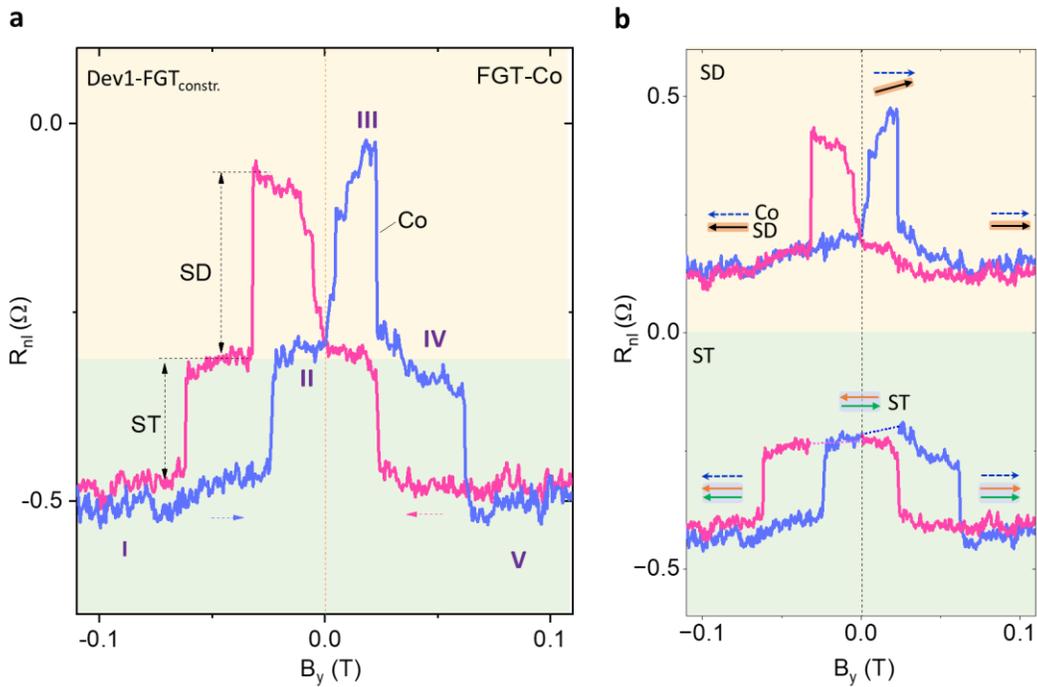

*Figure 2. Electrical detection of spin textures in van der Waals magnetic spin-valve. a. Measured spin-texture spin valve (ST-SV) signal using constricted FGT$_{constr.}$ as the injector and Co as the detector at room temperature in Dev 1. Stage I-V corresponds to different spin-texture statuses as a function of $B_y$. SD and ST represent the signal*

*components corresponding to single domain and spin texture.* ***b***. *The decomposed components, ST and SD, from the raw data in a) and the magnetic status at each stage. The black arrow indicates the SD component. The green and orange arrows represent the effective sub-domain structures of the ST component projection along the y-axis. Dashed arrows show the B field sweep directions.*

**Spin-texture spin valve device with a constriction in $Fe_5GeTe_2$**

First, we schematically explain the expected differences between spin texture and conventional spin valve signals as a function of the external magnetic field (Fig. 1c and Fig. 1d). The FGT with a **constriction** should have a region with spin textures. Compared to the conventional single domain (SD) spin valve signal with the shifted stage signals due to magnetic hysteresis effect of FGT and Co (Fig. 1d), the spin-texture (ST) spin valve signal is expected to have an overlapped stage signal (Fig. 1c). In ST spin valve signal, a sharp switch from low resistance to a high-resistance state occurs before the magnetic field sweep crossing the zero point due to spin texture's lower energy status with compensated magnetic moments. A switch from the high-resistance state to the low-resistance state occurs as the external field increases further, realigning the spin texture.

Experimentally, the constricted FGT spin texture spin valve data is measured at room temperature as a function of $B_y$, as presented in Fig. 2a. A multi-state switching is observed for both forward and backward $B_y$ magnetic fields sweep, due to the mixture of ST and SD spin valve signal. Figure 2b shows the decomposed SD and ST component signals with the corresponding magnetic moment status in the constricted FGT. First, we explain the forward magnetic field sweep curve (blue color), where at both large negative and large positive field ranges (state I and V), all magnetic domains are reoriented to align with the field into a single magnetic moment in parallel with the Co detector, leading to low resistance states of the spin valve signal. However, at low fields (stage II), the spin texture state of FGT remains more energetically favorable around zero fields, thereby resulting in a near-zero net magnetization and, equivalently, an intermediate resistance state (stage II) in the spin valve signal. After crossing the zero field from $-B_y$ to $+B_y$, the transition from stage II to III is attributed to the conventional magnetic domain switching of FGT. To be noted, the slow decrease of stage IV is due to the canted domain realignment to the y-axis. This is in agreement with our previous work with a conventional $FGT_{ref.}$/graphene heterostructure [23], indicating the coexistence of spin textures and canted magnetic domains in the constricted FGT at room temperature. Therefore, strictly speaking, stages IV and II are not due to the same magnetic domain state, though they have almost the same signal magnitude. A further increase in the magnetic field (around 60 mT) breaks the spin texture and aligns all of them to form a single domain again (stage V). When the magnetic field sweeps back, a similar signal can still be observed, proving the robustness of the spin texture spin valve signals.

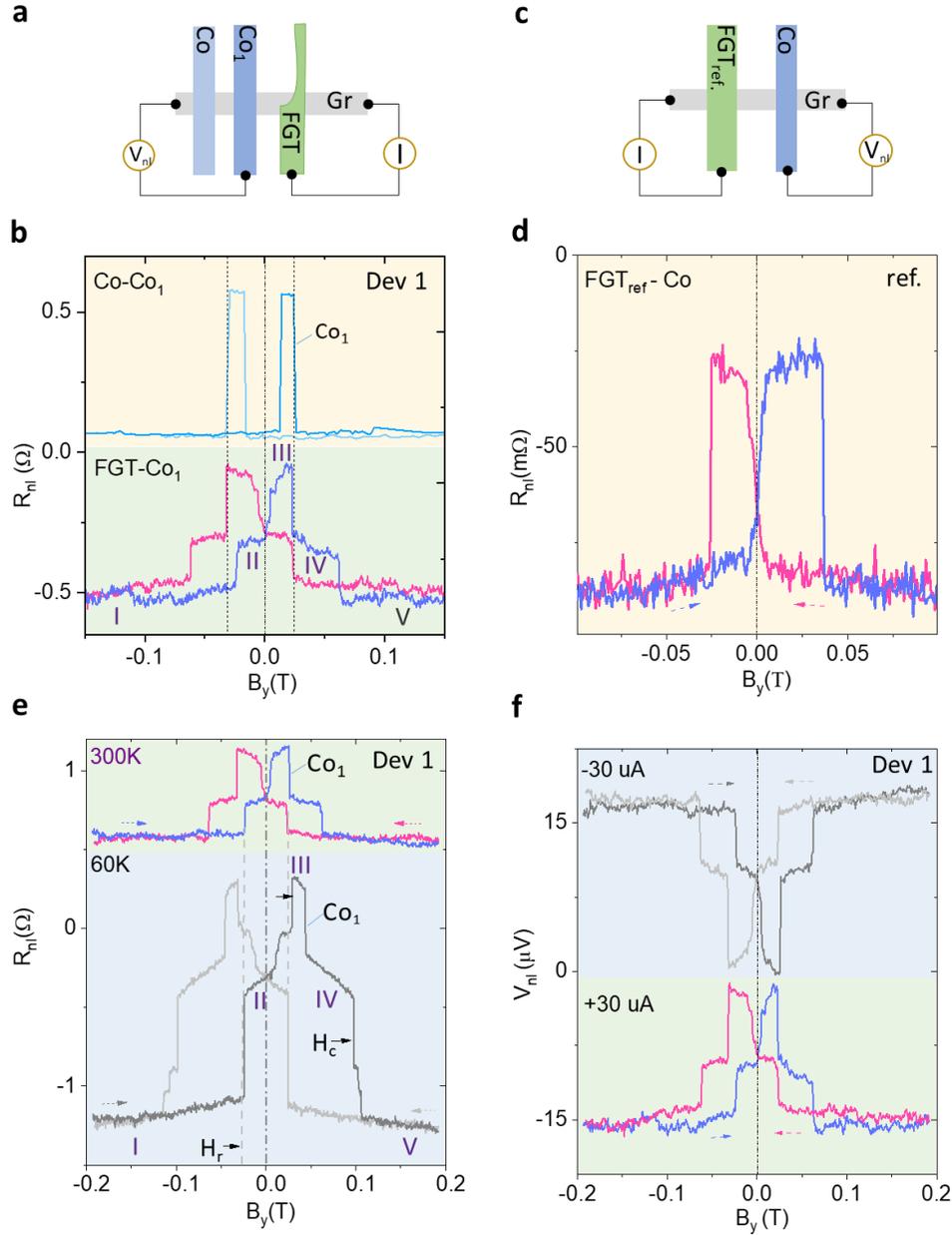

**Figure 3. Comparison of vdW magnet spin-texture spin-valve with conventional spin-valve signals. a, b.** Spin valve measurements schematics and signals for constricted FGT-Co$_1$ and reference Co-Co$_1$ device in Dev 1. **c, d.** Representative spin valve measurement geometry and signal as reference FGT$_{ref}$ - Co device. The reference contact FGT$_{ref}$ device has the stripe-shaped FGT without constriction or notch. **e.** Temperature-dependent spin-texture spin-valve signals at 60 K and 300 K of Dev 1 at a bias current of +30 μA with a coercive field ($H_c$) and recover field ($H_r$). **f.** ST spin-valve signal for opposite current polarity (+/- 30 μA) of Dev 1, resulting in the reversal of the signal.

**Control experiments and comparison with conventional single-domain spin valve devices**

In comparison to conventional Co-Co and reference $FGT_{ref}$-Co spin valves, we can distinguish the conventional SD and ST components in the spin valve signals more clearly. Figure 3a and 3b show the constricted FGT-$Co_1$ and Co-$Co_1$ spin valve configurations and signals, where both share the same $Co_1$ contact as the detection electrode in Dev 1. The same coercive field of $Co_1$ contact suggests that the separation of stage III and stage IV is merely due to the Co magnetization reversing (Fig. 3b). That is, stage III and stage IV are from the same type of magnetic domain in FGT, i.e., a continuous realignment from its tilt easy axis to $+B_y$ [23]. A further repeating observation of the spin texture spin valve signal with different Co detectors and measurement geometry confirms the robustness of the spin injection of ST and SD components in such constricted FGT (Supplementary Fig. S1a and S1b). Moreover, a comparison of the constricted-FGT and the reference $FGT_{ref}$-Co spin valve signal (Fig. 3c and Fig. 3d) shows that the constriction in FGT is necessary for nucleation of spin textures.

Low-temperature spin valve measurements were also performed for Dev 1. In Fig. 3e, the low-temperature (60 K) spin-valve measurement is compared with the room-temperature (300 K) signal. The magnitude of the spin valve signal is significantly amplified at lower temperatures, and the coercive fields ($H_c$) are shifted to higher values. Finer switching features, evident at both low and high fields, become apparent, which can be attributed to the more subtle spin textures appearing in FGT at lower temperatures[59]. For the SD contribution, the coercive field (stage II to stage III) has increased as expected due to enhanced magnetic anisotropy at lower temperatures, similar to that of the Co contact. Noticeably, we also observe a non-uniform shift in the coercive fields of the ST component. As highlighted with the dashed lines and arrows in Fig. 3e, the 'recover field' of the ST component, i.e., transitions from stage I to stage II, does not change with temperature. However, the coercive field (stage IV to stage V) has increased. This observation is consistent during both the forward-sweep and backward-sweep of the magnetic fields. It is also seen with a negative bias current (see Supplementary Fig. S1d). This behavior of the spin texture is quite different from that of conventional FM domains like Co and SD counterparts in FGT. This can be due to the pinning effect of the spin textures caused by the constriction. Figure 3f shows the effect of different bias polarities, where the sign of the voltage signal reverses but the signal magnitude stays comparable (see detailed bias dependence in Supplementary Fig. S1d).

**Spin-texture spin valve device with a notch in Fe$_5$GeTe$_2$**

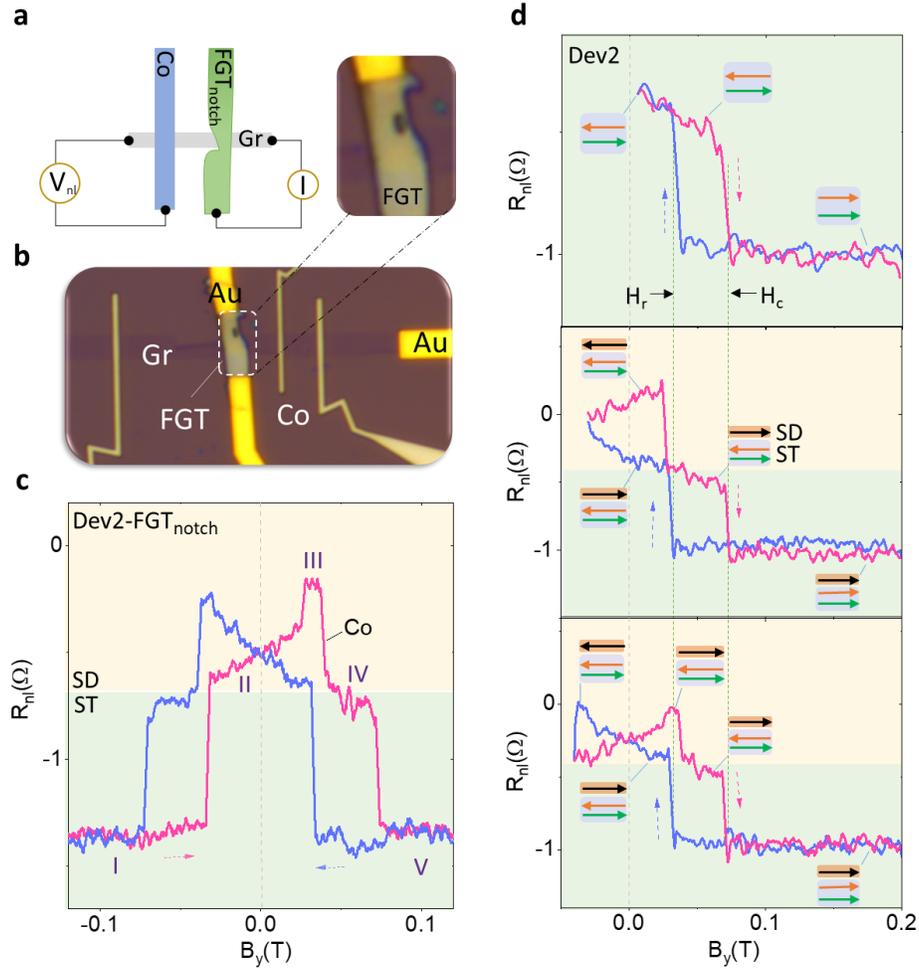

*Figure 4. **Spin-textures spin valve signal with a notch in Fe$_5$GeTe$_2$ of Dev 2.** **a, b.** Schematics and optical image of Dev 2 with a notch on FGT. **c.** Measured spin-texture spin-valve signals with a notched FGT. SD and ST components are marked with different background colors. **d.** Minor loop measurements of the spin valve signal in Dev 2. Green (cyan) color curves are the forward (backward) sweep, respectively. Insets show the corresponding domain structures at the critical field range. The black arrow indicates the SD component. The green and orange arrows represent the effective domain structures of the ST component projection along the y-axis.*

Next, we investigated a spin-texture spin-valve device with a notch in the FGT electrode (Dev 2) to investigate the pinning effect of the novel magnetic spin textures with different geometry[53,55] (Fig. 4a and Fig. 4b), which shows similar multi-state switches (stages I-V) with the coexistence of ST and SD states (Fig. 4c). A further detailed bias dependence of the SD and ST components shows a linear trend as a function of the bias (see Supplementary Fig. S1d), proving the linear regime spin injection of both SD and ST components at the small bias range. Detailed investigation of the minor loop measurement (Fig. 4d) in Dev 2 shows the hysteresis effect of the ST and SD component-induced spin valve signals [60]. The top panel in Fig. 4d shows that the ST component has a large hysteresis effect. Noticeably, it always displays the same 'alignment' $H_c$ and $H_r$, respectively, regardless of the magnetic field sweeping history. The middle panel of Fig. 4d proves the hysteresis effect of the conventional ferromagnetism (SD component) of FGT; while the bottom one confirms the hysteresis effect for both SD and ST components. This hysteresis effect in both SD and ST components exhibits a similar magnetic nature in response to the external field. Thus, the spin texture-related spin valve signals observed in different well-defined constricted geometries are strong evidence to prove the robustness of the nonlocal detection method and suggest that the spin textures can be well pinned at the FGT constrictions or notches.

**Hanle spin precession measurements in $Fe_5GeTe_2$ spin-texture spin-valve devices**

Hanle precession measurements are conducted to analyze the spin components with an out-of-plane magnetic field (zHanle) to assess in-plane spin polarizations ($S_x$ and $S_y$). While xHanle measurements were carried out with a magnetic field applied along the graphene channel (x-axis), resulting in the precession of spins in the yz-plane and providing information on the injection of $S_y$ and $S_z$ polarizations. Figure 5a and 5b illustrate the zHanle signal (constricted-FGT Device1) in comparison to the $B_y$ spin valve signal, displaying the spin states from the components of the constricted FGT near zero field. Interestingly, by comparing it with the spin valve signal ($B_y$), the Hanle signal ($B_z$) indicates that the main contribution arises from spin injection from the SD part of the FGT. Furthermore, both symmetric and antisymmetric Hanle components extracted from the Hanle signal suggest the presence of $S_y$ and $S_x$ spin polarizations (Supplementary Fig. S2a), aligning with the canted magnetization of FGT[61]. In contrast, the effective contributions from the ST components in the constricted FGT at low B fields remain anti-parallel, indicating almost zero net magnetization and thus no contribution to the Hanle signal.

More carefully controlled z(x)-Hanle experiments were performed with the pre-initiation of the parallel (P) and antiparallel (AP) magnetic alignments of the Co and the SD component of FGT (as schematically shown in Fig. 5c and Fig. 5e). In the zHanle signals of notched-FGT (Device 2), it is confirmed that the SD contribution dominates for both P and AP configurations. Furthermore, the ST contribution to the zHanle signal can be ruled out by comparing the Hanle signals (P and AP) with the reference ($B_y$) spin valve signal magnitude (Fig. 5d). The extracted symmetric and antisymmetric signals from the averaged signal $R_{nl}=[R_{nl}(AP)-R_{nl}(P)]/2$ indicate that both $S_y$ and $S_x$ spin polarization exist simultaneously in the SD component of the notched FGT as well (see Supplementary Fig. S2c).

Furthermore, by comparing the xHanle signals (AP and P) with the reference spin valve signal ($B_y$) in Fig. 4f, we can confirm that the contribution is mainly from the SD component at a lower field as well. However, at the high field range of the xHanle signals, the switching behavior for both P and AP configurations behaves similarly to the reference ($B_y$) spin valve signal with a larger coercive field, which suggests the realignment for both SD and ST components (see more detailed analysis in Supplementary Note 1). By averaging the xHanle P and AP signals, we can remove the non-spin-precession-related contributions [23]. The extracted symmetric and anti-symmetric components of the averaged signal are trivial at a large field range (Supplementary Fig. S2e). In agreement with the zHanle signal, the xHanle spin precession is dominated by the mainly symmetric component, showing that the majority of the spins injected are oriented along y, and with a finite $S_z$ spin population. Such Hanle measurements further confirm the existence of magnetic spin textures and the canted magnetism of the SD component in FGT.

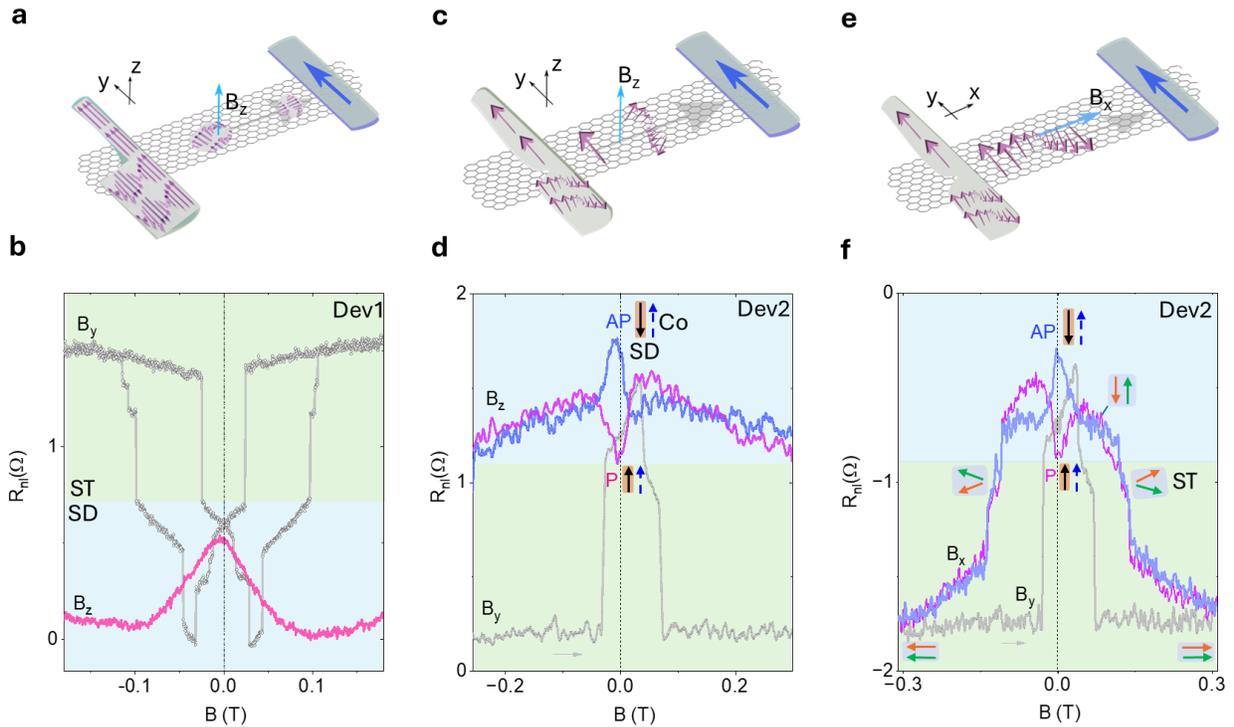

*Figure 5. Hanle spin precession measurements of spin-texture spin valve devices. a, b. Measurement geometry and comparison between the nonlocal zHanle signal and the $B_y$ spin valve signal of Dev 1. c, e. Schematics of the Measurement of the z(x)-Hanle measurements in the notched FGT-Co spin valve. d, f. z(x)-Hanle signals with parallel*

*(P) and anti-parallel (AP) magnetic moment configurations of Co and SD of notched FGT on Dev 2. Reference spin valve signals with the magnetic field $B_y$ as a comparison. For comparison, no shift along the y-axis was made.*

To gain further insights into the spin-texture-related magnetic domain rotation-induced switching behavior in the xHanle signals, detailed xy-plane and yz-plane angle-dependent measurements are presented in Supplementary Fig. S3a and Fig. S3b, showing that such domain-rotation-induced contribution at a larger field range is from the ST component. To be noted, such a metamagnetic transition under the influence of the $B_x$ field can only be observed when both the ST component and Co detector realign with the magnetic field $B_x$, as confirmed by the Co-Co xHanle signal (see Supplementary Fig. S3c and more detailed analysis of the angle-related evolution of the nonlocal signals in Supplementary Note 1). The detailed x(z)-Hanle signals with multiple spin polarizations confirm the tilted magnetization of the SD component in the constricted FGT [23]. The ST component is more clearly observed in the $B_y$ spin valve and magnetization rotation-induced Hanle signals. The observation of anomalous multi-level spin-valve switching and Hanle spin precession signals suggests that such a constriction/notch in FGT disrupts magnetic domains, enabling spin textures to persist alongside the conventional single magnetic domains (see more detailed discussion on the difference between SD and ST components in Supplementary Note 2).

Compared with $FGT_{ref}$ - graphene and Co-Co nonlocal spin valve, the constricted/notched FGT-Co spin valve shows a clear ST spin injection, besides the normal SD counterpart. Furthermore, the spin injection efficiency of SD and ST was also evaluated from the spin valve signal (Supplementary Note 3 and Table S1). It shows that both SD and ST have comparable spin injection efficiency $P_{ST}$ up to ~-20%, suggesting the ST component injection process behaves as efficiently as conventional SD signals. Moreover, a further analysis of the Hanle signals across the FGT/graphene heterostructure rules out the magnetic proximity effect [62–64] as the possible origin of the multi-level spin-valve switching behavior observed above (see Supplementary Note 4). For comparison with this unique spin texture of FGT, we also utilize well-studied material like Co to verify this electrical spin texture detection technique (see Supplementary Fig. S5). Experimentally, Co shows typical multiple domains in the notch and constriction geometry with signal stages I-III, while it only presents signal stages I-II in reference Co-Co striped structures. To be noted, all the switching happened after crossing zero magnetic field (Supplementary Fig. S5a-Fig. S5f).

**Conclusion and Outlook**

Our electrical detection of spin texture in $Fe_5GeTe_2$ reveals fundamental aspects of magnetism in van der Waals materials that distinguish them from conventional ferromagnets. As $Fe_5GeTe_2$ is known to exhibit stronger DMI than conventional ferromagnets, a simple notch and constriction can nucleate complex chiral spin configurations including skyrmions and stripe domains. We introduced an all-electrical method for detecting such spin textures in $Fe_5GeTe_2$ using pure spin transport in a graphene-based lateral spin-valve device. By engineering constrictions or notches in $Fe_5GeTe_2$, we created robust spin textures, which were injected into and electrically detected through a graphene channel at room temperature. This work marks the first demonstration of graphene as a spin transport channel for sensing spin textures in vdW magnets, enabling multi-level spin valve switching and Hanle precession signals without reliance on conventionally used microscopy techniques. The methodology established here for FGT can be extended to other 2D magnets with varying DMI strengths to understand how DMI emerge from crystal symmetry breaking in reduced dimensions. Such studies could guide the design of voltage-tunable spin textures in 2D heterostructures, where electrostatic gating can modulate DMI and stabilize different topological states on demand. The ability to detect complex spin textures in a simple 2D heterostructure by all-electrical methods can pave the way for the design of programmable spin logic elements and neuromorphic architectures[66]. Our approach stands apart from existing magnetic tunnel junction[67] or racetrack memory[68] techniques developed for conventional magnetic layers, offering a simpler and fully 2D architecture with direct electrical access to spin textures. Looking forward, this methodology can be extended to investigate dynamics of nanoscale skyrmions, other topological textures and could form the basis of neuromorphic computing architectures and unconventional memory elements that capitalize on the rich spin landscape of 2D quantum materials.

**Methods**

**Fabrication of devices and electrical measurements**

The CVD graphene used for devices was transferred onto a 4-inch n[++]Si substrate with a 285 nm $SiO_2$ layer. CVD graphene channels were prepared first using electron beam lithography (EBL) and oxygen plasma etching. The FGT flakes (20-50 nm in thickness) were exfoliated from single crystals grown by physical vapor transport (PVT) method (from HqGraphene) and dry transferred onto monolayer CVD graphene inside a $N_2$ glovebox. Exfoliation of FGT and preparation of their heterostructures with CVD graphene were carried out inside a glove box under $N_2$ atmosphere to obtain a clean interface. For device fabrication, nonmagnetic and magnetic contacts were prepared using multiple EBL processes and electron beam evaporation of metals. Nonmagnetic Au/Ti contacts were first prepared on FGT flakes with a 10-second low-energy Ar cleaning of the surface at a glancing angle. Next, nonmagnetic Au/Ti contacts were prepared on graphene for reference electrodes using EBL and the lift-off process. Ferromagnetic contacts (Co/TiO$_2$) on graphene were prepared using EBL, electron beam evaporation, and a lift-off process. For Co/TiO$_2$ contacts, a two-step deposition and oxidation process was adopted: 0.4 nm Ti was deposited, followed by a 10 Torr O$_2$ oxidation for 10 min each, followed by 60 nm Co deposition. Measurements were performed at room temperature with a magnetic field up to 0.6 Tesla and a sample rotation stage under vacuum conditions. Electronic measurements were carried out using a current source Keithley 6221, and a nanovoltmeter 2182A.

**Acknowledgments**

Authors acknowledge funding from European Innovation Council (EIC) project 2DSPIN-TECH (No. 101135853), 2D TECH VINNOVA competence center (No. 2019-00068), European Commission (EU) Graphene Flagship, Wallenberg Initiative Materials Science for Sustainability (WISE) funded by the Knut and Alice Wallenberg Foundation, EU Graphene Flagship (Core 3, No. 881603), Swedish Research Council (VR) grant (No. 2021–04821, No. 2018-07046), FLAG-ERA project 2DSOTECH (VR No. 2021-05925) and MagicTune, Carl Tryggers foundation, Graphene Center, Chalmers-Max IV collaboration grant, VR Sweden-India collaboration grant, Areas of Advance (AoA) Nano, AoA Materials Science and AoA Energy programs at Chalmers University of Technology and Nanofabrication laboratory MyFab at Chalmers University of Technology.

**Data availability**

The data that support the findings of this study are available from the corresponding authors on a reasonable request.

**Author Contributions**

B.Z., S.P.D. conceived the idea and designed the experiments. B.Z. and R.N fabricated and characterized the devices. B.Z., R.N., L.S., and S.P.D. analyzed and interpreted the experimental data, compiled the figures, and wrote the manuscript with inputs from all co-authors. S.P.D. supervised the research project.

**Competing interests**

The authors declare no competing interests.